\documentclass[superscriptaddress,twocolumn,showpacs,preprintnumbers,prb]{revtex4}
\usepackage{graphicx}
\usepackage{times}

\usepackage{color}

\begin{document}

\title{Theoretical Study of Half-Doped Models for Manganites:\\
Fragility of the CE Phase with Disorder, Two Types of 
Colossal Magnetoresistances, \\
and Charge-Ordered States for Electron-Doped Materials}

\author{H. Aliaga}
\affiliation{National High Magnetic Field Lab and Department of Physics, 
Florida State University, Tallahassee, FL 32310.}
\author{D. Magnoux}
\affiliation{Laboratoire de Physique
Theorique CNRS-FRE2603, Universit\'{e} Paul Sabatier, F-31062 Toulouse, France.}
\author{A. Moreo}
\affiliation{National High Magnetic Field Lab and Department of Physics, 
Florida State University, Tallahassee, FL 32310.}
\author{D. Poilblanc}
\affiliation{Laboratoire de Physique
Theorique CNRS-FRE2603, Universit\'{e} Paul Sabatier, F-31062 Toulouse, France.}
\author{S. Yunoki}
\affiliation{International School for Advanced Studies (SISSA), via Beirut 4, 34014 Trieste, Italy.} 
\author{E. Dagotto}
\affiliation{National High Magnetic Field Lab and Department of Physics, 
Florida State University, Tallahassee, FL 32310.}

\date{\today}

\begin{abstract}

A comprehensive analysis of half-doped manganites is presented
using Monte Carlo simulations applied to the double-exchange model
with cooperative Jahn-Teller lattice distortions in two dimensions. 
A variety of novel results are reported. In
particular: {\it (i)} The phase diagram is established in the $\lambda$-$J_{\rm AF}$
plane, with $\lambda$ the electron-phonon coupling and $J_{\rm AF}$
the antiferromagnetic exchange between classical $t_{\rm 2g}$ spins.
The results include standard phases, such as the CE-insulating and FM-metallic
regimes, but they also include novel states, such as a
ferromagnetic charge-ordered (CO) orbital-ordered phase originally
predicted by Hotta {\it et al.}. This state is compatible with
recent experimental results by Loudon {\it et al.} 
{\it (ii)} For realistic couplings, it was observed that 
the charge disproportionation $\delta$
of the CO phase is far from the widely accepted extreme limit
$\delta$=0.5 of a  3+/4+ charge separation. A far smaller
$\delta$ appears more realistic,
in agreement with recent experiments by Garcia {\it et al.} and
Daoud-Aladine {\it et al.}  
{\it (iii)} Colossal magnetoresistance (CMR) effects are
found in calculations of cluster resistances using the Landauer formalism.
This occurs near the ubiquitous first-order phase transitions between
the insulating and metallic states. The present result 
reinforces the previous conjecture
that CMR phenomenology exists in two forms: the low-temperature CMR
addressed here and the more standard CMR above the Curie temperature.
{\it (iv)} The CE-state is found to be {\it very sensitive to disorder} since
 its long-range order rapidly disappears when quenched-disorder
is introduced, contrary to the FM state which is more robust. This is
also in qualitative agreement with recent experiments by Akahoshi {\it et al.}
and Nakajima {\it et al.}
{\it (v)} The phase diagram in the half-doped {\it electron doping} regime
is briefly discussed as well. A charge-ordered
state is found which is the analog of the $x$=0.5 CE phase. It
contains a 3+/2+ charge arrangement at large
$\lambda$. Numerical results suggest that 
an approximate symmetry exists between the
hole- and electron-doped systems in the large Hund coupling
limit.

\end{abstract}


\maketitle

\section{Introduction}

\begin{figure}[h]
\includegraphics[width=6cm]{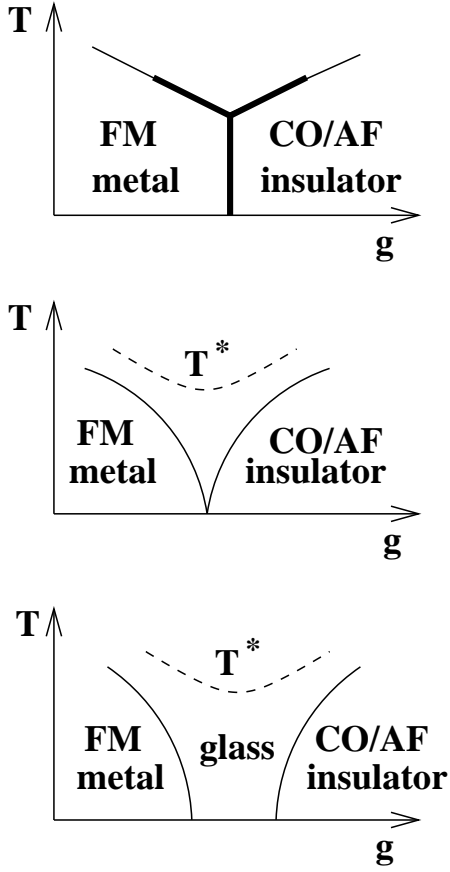}
\caption{ {\it Top} General phase diagram of two competing phases in the absence of
quenched-disorder (or when this disorder is very weak). Thick (thin) lines
denote first (second) order transitions. Shown is a tricritical case, but it
could be bicritical or tetracritical as well. $g$ is some parameter needed
to change from one phase to the other. \textit{Middle} With increasing
disorder, the temperature range with first-order transitions separating the
ordered states is reduced, and eventually for a \textit{fine-tuned} value of
the disorder the resulting phase diagram contains a quantum critical point.
In this context, this should be a rare occurrence. \textit{Bottom} In the
limit of substantial quenched
disorder, a window without any long-range order
opens at low temperature between the ordered phases.
This disordered state has glassy characteristics and it is composed of
coexisting clusters of both phases. The size of the coexisting islands can
be regulated by the disorder strength and range, and by the proximity to the
original first-order transition. For more details 
see Refs.~\onlinecite{burgy,book}. 
The new scale $T^{*}$ discussed in the text-- remnant of the
clean-limit transition -- is also shown. }
\label{fig:tg3}
\end{figure}

Manganites are currently attracting considerable attention
mainly due to the presence of the colossal magnetoresistance effect
in magnetotransport measurements\cite{reviews,book}. In addition, these materials
have a complex phase diagram with a plethora of ordered phases,
a typical characteristic of correlated electron systems. Many experimental
and theoretical investigations
have unveiled the inhomogeneous character of the states of relevance
to explain the CMR phenomenon, with a competition between ferromagnetic
and antiferromagnetic states that induces coexistence of clusters,
typically with nanometer-scale sizes\cite{book}.
The rationalization of this
phenomenon, and concomitant explanation of the CMR effect, is based on the
first-order transitions that separate the metallic and insulating phases
in the clean limit (i.e. without disorder)\cite{book,reviews}. 
The first-order character
of the transition is caused by the different magnetic and charge
orders of the competing states. The clean-limit 
phase diagram  is illustrated in 
Fig.~\ref{fig:tg3} (upper panel). When quenched
disorder is introduced in the coupling or density that is modified
to change from one phase to the other, the temperature where the N\'eel
and Curie temperatures meet is reduced in value and eventually reaches
zero as in Fig.~\ref{fig:tg3} (middle panel). Upon further increase 
of the disorder strength, a spin
disordered region appears at low temperatures, with a glassy behavior
involving coexisting clusters (Fig.~\ref{fig:tg3} (lower panel)).
Simulations by Burgy {\it et al.} \cite{burgy} have shown that the
clustered state between the Curie temperature and the clean-limit
critical temperature ($T^*$), with preformed ferromagnetic regions of random orientations,
has a huge magnetoresistance since small fields can easily align
the moments of the ferromagnetic islands, leading 
to a percolative conductor in agreement with experiments\cite{fath}.
The quenched disorder simply triggers the stabilization of the cluster
formation, but phase competition is the main driving force 
of the mixed state.
It has been speculated\cite{burgy} that similar phenomena should occur in a variety
of materials, including, for example, 
the high temperature superconductors where nanoclusters have been found\cite{pan}.

In spite of the phenomenological
success of the previous calculations discussed above, there are
many issues that must be further 
investigated in the manganite context to reach
a reasonable understanding of these materials (for a recent discussion
on open issues see Ref.~\onlinecite{open}). It is the purpose
of this paper to contribute to manganite studies by addressing
a variety of open topics, using computational techniques applied to 
realistic models for Mn-oxides. In particular, here the phase
diagram at hole-density $x$=0.5 is presented for the case of
{\it cooperative} Jahn-Teller phonons in interaction with carriers.
Previous studies mainly focused on non-cooperative phonons\cite{previous}.
The phase diagram reported below in Fig.~\ref{fig:figure1} has interesting features. For
example, it contains the realistic FM charge-disordered and 
CE charge-ordered phases well-known from experiments.
In addition, further evidence is provided that the transition from
the FM to the AF state is first-order at low temperatures. This
first-order property allows us to obtain a huge magnetoresistance
effect even working on small clusters, since small magnetic fields
can unbalance the ground state from antiferro to ferro, in the vicinity
of the competition region. For the calculation, techniques borrowed
from mesoscopic calculations using the Landauer formalism are used.
The shape of the phase diagram with increasing temperature is in
good qualitative agreement with recent experiments that established
the bicritical form of
the $x$=0.5 phase diagram\cite{Tomioka}. Moreover, a novel phase is found with both
charge and ferromagnetic order, in agreement with previous
calculations\cite{hotta-feiguin,previous} 
and also with recent experiments by Loudon {\it et al.}\cite{mathur} 
Another feature
of the results found here that is in good agreement with recent
experiments (to be discussed below)
is the value of the charge disproportionation. The widely accepted
view of the extreme 3+/4+ separation needs to be revisited. A much
milder charge separation appears more realistic.

The $x$=0.5 phase diagram of the realistic model studied here
is also analyzed in the presence
of quenched-disorder, to test the results of phenomenological models,
such as those shown in Fig.~\ref{fig:tg3}. Overall, good
agreement with previous studies
is found, but surprises are  also reported such as the 
strong sensitivity of the CE phase to quenched disorder. This
behavior is also in good agreement with recent experimental results by 
Akahoshi {\it et al.}\cite{akahoshi} and Nakajima {\it et al.}\cite{ueda}, 
where an asymmetry 
between the behavior of the FM- and CE-phases was observed when
disordered was introduced in some materials with $x$=0.5.  
Finally, since {\it electron} doping of manganites
has also been experimentally actively pursued \cite{electron1},
the model studied here is analyzed for the case of electron
doping as well, with an electronic density per site equal to 1.5. 
A state quite similar to the CE phase is observed, which could
be found in future experiments. 

The overall conclusion is that theoretical studies of realistic models
for manganites using unbiased techniques 
are unveiling a remarkable
qualitative agreement with experiments.
While previous phenomenological descriptions of manganites are reasonable
approximations for the understanding of the 
puzzling CMR phenomenon\cite{book}, many other features observed 
in the present studies
still need further investigations, such as the strong
sensitivity to disorder of the CE state.

\section{Method and Definitions}

The Hamiltonian studied in this paper is 
\begin{eqnarray}
H &=&-\sum_{\bf{ia}\gamma \gamma ^{\prime }\sigma }t_{\gamma \gamma
^{\prime }}^{\bf{a}}(d_{\bf{i}\gamma \sigma }^{\dag }d_{\bf{i+a}%
\gamma ^{\prime }\sigma }+h.c.)-J_{\rm{H}}\sum_{\bf{i}}\bf{s}_{\bf{%
i}}\cdot \bf{S}_{\bf{i}}  \nonumber \\
&+&J_{\rm{AF}} \sum_{\langle \bf{i,j} \rangle } \bf{S}_{\bf{i}}\cdot \bf{S}_{\bf{j}}  
+\lambda \sum_{\bf{i}}(\it{Q}_{\rm{1}\bf{i}}\rho
_{\bf{i}}+Q_{\rm{2}\bf{i}}\tau _{\rm{x}\bf{i}}+Q_{\rm{3}\bf{i}}\tau _{\rm{z}\bf{i}})  \nonumber \\
&+&(1/2)\sum_{\bf{i}}(\beta Q_{1\bf{i}}^{2}+Q_{2\bf{i}}^{2}+Q_{3%
\bf{i}}^{2}),
\end{eqnarray}
where $d_{\bf{i}\rm{a}\sigma }$ ($d_{\bf{i}\rm{b}\sigma }$)
annihilates an $e_{\rm{g}}$-electron with spin $\sigma $ in the $%
d_{x^{2}-y^{2}}$ ($d_{3z^{2}-r^{2}}$) orbital at site $\bf{i}$, and $%
\bf{a}$ is the vector connecting nearest-neighbor (NN) sites. The first
term is the NN hopping of $e_{\rm{g}}$ electrons with amplitude $%
t_{\gamma \gamma ^{\prime }}^{\bf{a}}$ between $\gamma $- and $\gamma
^{\prime }$-orbitals along the $\bf{a}$-direction: $t_{\rm{aa}}^{%
\bf{x}}$=$-\sqrt{3}t_{\rm{ab}}^{\bf{x}}$= $-\sqrt{3}t_{\rm{ba%
}}^{\bf{x}}$=$3t_{\rm{bb}}^{\bf{x}}$=$t$ for $\bf{a}$=$%
\bf{x}$, and $t_{\rm{aa}}^{\bf{y}}$=$\sqrt{3}t_{\rm{ab}}^{%
\bf{y}}$= $\sqrt{3}t_{\rm{ba}}^{\bf{y}}$=$3t_{\rm{bb}}^{%
\bf{y}}$=$t$ for $\bf{a}$=$\bf{y}$ (our study will be restricted 
to two-dimensional lattices).
Hereafter, $t$ is taken as the energy unit. In the second term, the Hund
coupling $J_{\rm{H}}$($>$0) links $e_{\rm{g}}$ electrons with spin $%
\bf{s}_{\bf{i}}$= $\sum_{\gamma \alpha \beta }d_{\bf{i}\gamma
\alpha }^{\dag } \bf{{\sigma }_{\alpha \beta }}$ $d_{i\gamma \beta }$ ($%
\bf{\sigma }$=Pauli matrices) with the localized $t_{\rm{2g}}$-spin $%
\bf{S}_{\bf{i}}$, assumed classical with $|\bf{S}_{\bf{i}}|$%
=1. $J_{\rm{H}}$ is here considered as infinite or very large. The third
term is the AFM coupling $J_{\rm{AF}}$ between NN $t_{\rm{2g}}$
spins. The fourth term couples $e_{\rm{g}}$ electrons and MnO$_{6}$
octahedra distortions, $\lambda $ is a dimensionless
coupling constant, $Q_{1\bf{i}}$ is the breathing-mode distortion, $Q_{2%
\bf{i}}$ and $Q_{3\bf{i}}$ are, respectively, $(x^{2}$$-$$y^{2})$-
and $(3z^{2}$$-$$r^{2})$-type JT-mode distortions, $\rho _{\bf{i}}$= $%
\sum_{\gamma ,\sigma }d_{\bf{i}\gamma \sigma }^{\dag }d_{\bf{i}%
\gamma \sigma }$, $\tau _{\rm{x}\bf{i}}$= $\sum_{\sigma }(d_{\bf{%
i}\rm{a}\sigma }^{\dag }d_{\bf{i}\rm{b}\sigma }$ +$d_{\bf{i}%
\rm{b}\sigma }^{\dag }d_{\bf{i}\rm{a}\sigma })$, and $\tau _{%
\rm{z}\bf{i}}$= $\sum_{\sigma }(d_{\bf{i}a\sigma }^{\dag }d_{%
\bf{i}a\sigma }$ $-$$d_{\bf{i}b\sigma }^{\dag }d_{\bf{i}b\sigma
})$. The fifth term is the usual quadratic potential for adiabatic
distortions and $\beta $ is the spring-constants ratio for breathing- and
JT-modes. In actual manganites, $\beta $$\approx $2 (see Ref.~\onlinecite{hotta}), 
and this is the value, unless something different is stated, we will consider throughout 
this paper. In undoped manganites, all oxygens are shared by adjacent 
MnO$_{6}$ octahedra and the distortions 
are not independent, suggesting that the {\it cooperative} 
effects are very important. This observation is likely valid even 
at finite hole densities, since experiments show the presence 
of orbital ordering  in the half-doped regime.
To consider this cooperation, here
oxygen ion displacements, denoted by $u_{\bf i}^{x}$ and $u_{\bf i}^{y}$, 
are directly optimized \cite{oxygen}.

It has been clarified in previous literature (e.g., Ref.~\onlinecite{book}) 
that a large Hund coupling as
used here suppresses double occupancy of the same orbital, and 
in this respect behaves as a Hubbard $U$ interaction. In addition,
a robust coupling $\lambda$ also acts as a $U^{'}$ repulsion between
electrons in the same site but different orbitals. As a consequence,
it is a reasonable approximation to neglect the
 Coulombic interactions in
the problem, which simplifies enormously the computational effort.

The model will be analyzed primarily using a classical Monte Carlo (MC)
procedure for the localized spins and phonons, in conjunction with exact
diagonalization of the conduction electron system. This last part of the process
corresponds to the solution of the single-electron problem with hoppings determined 
by the localized spin configuration. 
The resulting energy levels are then filled with the number
of electrons to be studied, namely the simulations are carried out in the 
canonical ensemble (see Appendix). 
However, simulations directly in the grand canonical ensemble
varying chemical potentials were also carried out.
In the present work, all the calculations were made on a 2D 4$\times$4 
cluster, mainly 
with filling $x$=1/2, as explained in the introduction. Currently, 
it is not possible to comprehensively study
larger systems, unless in special cases, due
to the considerable CPU time that the diagonalization process needs, and
the large number of parameters that must be varied to fully explore
the phase diagram. In spite of the size limitation of our effort, 
the results reported here have a clear
physical interpretation and size effects appear to be mild for
the quantities that were investigated.

Updates of the spin and phononic $\{\theta _{\bf i},\phi _{\bf i},u_{\bf i}^x,u_{\bf i}^y\}$ 
configurations are accepted or rejected according to the 
Metropolis algorithm. The number of MC steps per site is typically taken as 
3000 for thermalization, with an additional 10000 for measurements. 
The simulations usually started with random states at high temperature, 
and then the temperature was decreased slowly, but at very low temperatures
some of the simulations started with ordered states to speed up the convergence.
The results of the simulations were analyzed using
three separate but related quantities: the spin structure factor 
\begin{equation}
S(\bf{k})=\sum_{i,j}\langle \bf{S}_{i}\bf{.S}_{j} \rangle e^{i\bf{k.}(\bf{%
r}_{i}-\bf{r}_{j})},  \label{essf}
\end{equation}
of the classical spins, the charge structure factor 
\begin{equation}
N(\bf{k})=\sum_{i,j} \langle \rho_{i} \rho_{j} \rangle e^{i\bf{k.}(\bf{r}_{i}-\bf{r}%
_{j})},  \label{ecdf}
\end{equation}
and the orbital structure factor defined as,  
\begin{equation}
T(\bf{k})=\sum_{i,j} \langle \bf{T}_{i}\bf{.T}_{j} \rangle e^{i\bf{k.}(\bf{%
r}_{i}-\bf{r}_{j})},  \label{esso}
\end{equation}
with $\bf{T}_{i}=(\tau _{\rm{x}\bf{i}},\tau _{\rm{y}%
\bf{i}},\tau _{\rm{z}\bf{i}})$ and $\tau _{\rm{y}\bf{i}}$%
= $i \sum_{\sigma } (d_{\bf{i}\rm{a}\sigma }^{\dag }d_{\bf{i}%
\rm{b}\sigma }$ -$d_{\bf{i}\rm{b}\sigma }^{\dag }d_{\bf{i}%
\rm{a}\sigma })$. Frequently, visual investigations of snapshots 
of the spin and phononic configurations at low
temperatures were useful to guide the intuition. Some dynamical properties
were also studied, such as the density-of-states. For details see Ref.~\onlinecite{book}.
More sophisticated techniques used to study the conductance of clusters
are described later in the paper.

\section{Low-Temperature Phase Diagram}

\begin{figure}[t!]
\includegraphics[width=8cm]{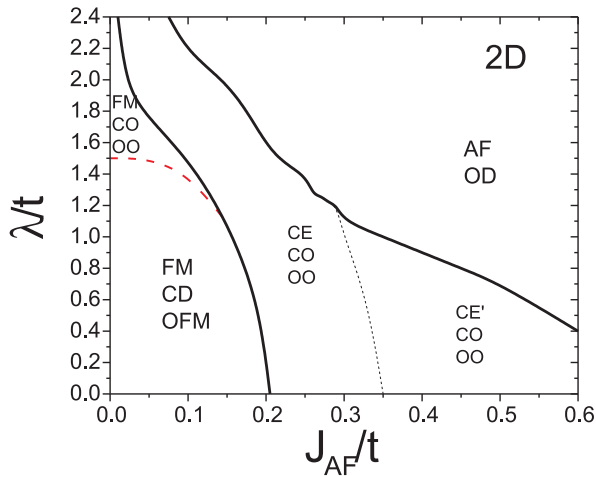}
\caption{Phase diagram of the half-doped 2D two-orbitals model with
cooperative Jahn-Teller phonons at $T/t$$\sim$0. The Hund coupling $J_{H}$ is
infinite. The phases were obtained analyzing the crossing of energies (as
explained in Fig.~\ref{fig:figure2}), 
and also $S({\bf k})$, $T({\bf k})$, and $N({\bf k})$ 
(see definitions in
text). The magnetic phases present are FM, CE, CE' and AF. The symbols CO(CD)
stand for charge-ordered (-disordered) phases. OFM corresponds to a
homogeneous state with dominant uniform $x^{2}-y^{2}$ orbital order. The
staggered orbital-ordered states, OO, have the standard ''CE'' pattern,
present in the three phases. The FM-CD-OFM phase is metallic, the rest of
the phases are insulating. Thick lines denote first-order transitions, the
dash-line is second order. The lines are smooth interpolations between a
finite but large number of points obtained numerically.}
\label{fig:figure1}
\end{figure}

\begin{figure}[h!]
\includegraphics[width=8cm]{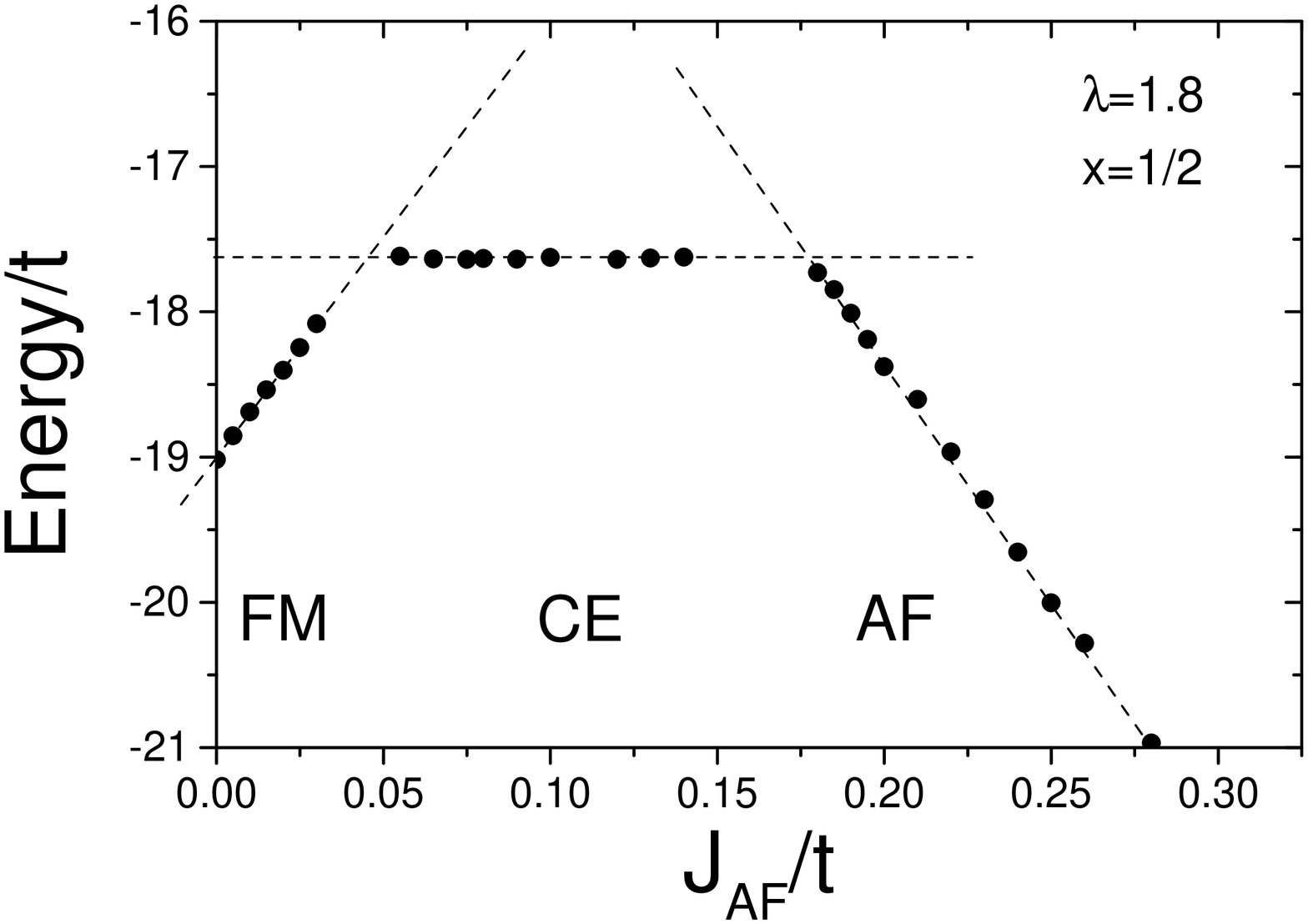}
\caption{MC energy vs. $J_{\rm AF}$ for $x$=1/2, $\lambda $=1.8, $T/t$$\sim$0.
This low-$T$ energy is calculated after starting with a random high-$T$
state and lowering $T$ slowly. For this value of $\lambda$, the three FM, CE 
and AF phases have OO with the $%
d_{3x^{2}-r^{2}}-d_{x^{2}-y^{2}}-d_{3y^{2}-r^{2}}$ usual alternation of
orbitals, distributed in zig-zags. The $x^{2}-y^{2}$ orbitals at the ``Mn$%
^{4+}$'' sites are populated at small and moderate values of $\lambda $.
The three phases also have checkerboard-type charge ordering. Near
the crossings of levels we have observed a notorious slowing down in the
convergence of the MC calculations. Lines are guides to the eye. Very
similar results were obtained at several values of $\lambda$ and were used to
construct Fig.~\ref{fig:figure1}}
\label{fig:figure2}
\end{figure}

The phase diagram of the double-exchange model with cooperative JT phonons
at $x$=0.5 is shown in Fig.~\ref{fig:figure1}. The properties of each phase
were analyzed through the spin, charge, and orbital correlations in real
and momentum space defined in the previous section. Varying $J_{\rm AF}$
three spin regimes were identified, similar to those reported in previous
investigations\cite{yunokiJT}. Typical results for the evolution of the energy
with $J_{\rm AF}$
are presented in Fig.~\ref{fig:figure2}, clearly showing the three states that compete
at fixed $\lambda$. Using this procedure, the phase diagram was constructed.
Note the clear first-order transitions separating the different phases, shown
in the figure as level crossings.

In agreement with well-known experimental
results for half-doped real manganites, a CE phase is stabilized in the
intermediate $J_{\rm AF}$ range. The actual correlations for this phase
will not be shown explicitly here since they have appeared in some previous
investigations\cite{yunokiJT} 
and our results agree very well with those results.
However, there are issues to remark not discussed before. For example,
note that the CE phase is present not only
at large $\lambda$ but also at small electron-phonon coupling. This last
result suggests that a nonzero $J_{\rm AF}$ is sufficient to stabilize the
zigzag chains characteristic of the CE region. 
Hotta {\it et al.}\cite{hottaCE} 
proposed that a band-insulator picture is needed to understand this phenomenon,
and the present results support this view. It is the geometry of the zigzag
chains that plays the key role 
in stabilizing the CE state. This result may have 
important consequences for experiments as discussed below. 
The dotted line represents a continuous transition between the standard CE 
phase and a novel canted CE' phase. To understand the characteristics of the 
new CE' state, note that the standard CE phase can also be interpreted 
as an arrangement of parallel AF zig-zag chains, displaced one lattice parameter 
in the x- and y-directions simultaneously. When the relative spin angle between 
the neighboring AF zig-zags is $\pi$, we have the normal CE phase. In the CE' 
regime this relative angle varies continuosly as $\lambda$ or $J_{\rm AF}$ are 
increased. Finally, when this spin angle between neighboring AF 
zig-zag chains is 0, we recover the normal AF phase.

At sufficiently large $J_{\rm AF}$ coupling, an antiferromagnetic spin
arrangement is found. This phase has not been observed yet experimentally,
but there is no reason to believe that the state cannot exist in some Mn oxide.
Within the accuracy of our study, 
this AF phase is believed to have no long-range order in the
orbital and charge degrees of freedom.
This corresponds to a nearly uniform arrangement of charge 
at small $\lambda$. However, at large electron-phonon coupling the situation
changes qualitatively. In this regime, a ``polaronic'' state seems to
be stabilized, where 
the 8 electrons present in the 16-sites cluster studied here
become trapped in random locations. If in this random distribution
some of the electrons are close to one another, local
orbital order similar to that of the $x$=0 limit was observed. Again, this
AF phase has not been observed experimentally yet, but the recent 
progress in the discovery of new phases (see next section) 
suggests that this remains a serious possibility.

The regime of small $J_{\rm AF}$ coupling is dominated by FM states. 
The large-$\lambda$ zone has a novel state which is discussed in the
next section. At intermediate and small couplings, the state that dominates
at low temperature has the characteristics of the FM metallic state
well-known to exist in three-dimensional Mn oxides. For instance, in
the charge sector there is no order. Regarding orbital order, the
correlations suggest the presence of a uniform arrangement.
This uniform orbital order at small $\lambda$ is not produced through
a spontaneous-symmetry-breaking process,
but it exists as a direct consequence of the form of the kinetic energy
in our two-dimensional simulations. To show this, the procedure is the
following. Consider the limit of Hund
coupling infinite as in our studies, and also $\lambda$=$J_{\rm AF}$=0. 
In this case, the Hamiltonian
can be diagonalized exactly in momentum space and the ground state
for the 4$\times$4 cluster can be exactly constructed. In this state,
the mean value of the number operator for the two orbitals of relevance can be
calculated as well. The results are not equal for the two orbitals, 
but there is an asymmetry
in favor of the $x^2$-$y^2$ orbital. Then, in this respect the uniform 
orbital-order in the state
is explicit in the model, and it is not induced by the JT phonons.
These conclusion were also checked on larger lattices, 
such as 20$\times$20, and in three-dimensional clusters.

\section{Novel FM/CO Phase}

\begin{figure}[h!]
\includegraphics[width=6cm]{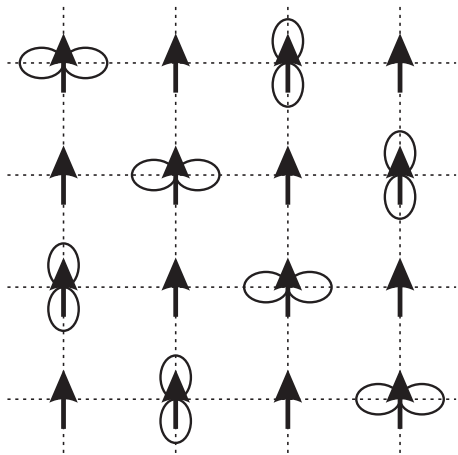}
\caption{Orbital, spin, and charge arrangement for the FM-CO-OO phase. 
The $3x^{2}-r^{2}$ and $3y^{2}-r^{2}$ orbitals represent the Mn$^{3+}$ ions, 
while the empty sites represents the Mn$^{4+}$ sites. This state was presented
for the first time in Ref.~\onlinecite{hotta-feiguin}. Evidence of a FM phase 
with checkerboard CO pattern has been found recently.\cite{mathur}}
\label{fig:figureFMCO}
\end{figure}

It is interesting to remark that the phase diagram in Fig.~\ref{fig:figure1}
contains not only the FM-metallic 
and CE-insulating phases -- well-established experimentally --
but other phases as well. For example, at large $J_{\rm AF}$ 
an AF
phase exists in a wide range, and this phase could be observed 
in future experimental investigations, as already discussed. 
Even more exciting is the case of the
ferromagnetic charge-ordered (FM/CO) phase 
in the upper-left corner of Fig.~\ref{fig:figure1}. This phase
was found in previous
investigations by Hotta {\it et al.}\cite{hotta-feiguin} using
cooperative phonons (and suggested by Yunoki {\it et al.}\cite{previous} 
as well, using non-cooperative phonons). It is quite
remarkable that 
recent experimental efforts by Loudon {\it et al.}\cite{mathur} 
have unveiled the presence of a FM insulating 
phase at half doping, compatible with the FM/CO state found in 
the Monte Carlo simulations.
For completeness, in Fig.~\ref{fig:figureFMCO} 
the orbital arrangement of the FM/CO phase
found in our simulations is shown, although it has been already
presented in Ref.\onlinecite{hotta-feiguin}. 
The orbital order corresponds to the same pattern of
$3x^2-r^2$/$3y^2-r^2$ orbitals of the CE-phase, but the spins are arranged
ferromagnetically, since $J_{\rm AF}$ is not large enough to induce the CE order.
Future experiments will clarify if indeed the theoretically
predicted FM/CO regime\cite{hotta-feiguin,previous} 
corresponds to the results reported in Ref.~\onlinecite{mathur}.
Note that the discovery of more phases than previously known is a recent interesting
trend, both in experiments and theory. 
In fact, at $x$=0 a novel E-phase has been observed in simulations by
Hotta {\it et al.}\cite{hottaEphase},
as well as in experiments by Kimura {\it et al.}\cite{newEphase}.  
Many surprises may still
be found in manganite investigations, even at the rather elementary
level of characterization of states,
since its strongly correlated character
creates a plethora of phases that are in strong competition.

\section{Phase Diagram Varying Temperature}

The properties of the model studied here were also investigated
at finite temperatures. In this case, the errors in the estimations
of the phase boundaries are larger
than at zero temperature. The reason is that at $T$=0, the first-order
transitions can clearly be established through level crossings, which exist
even in small systems. However, the continuous transitions that are found
at finite $T$ can only be roughly located on small systems based on 
spin, charge, and orbital correlations at the largest distance available
in the clusters studied. In spite of this problem, the results shown here
are sufficiently accurate to understand the main trends in the phase diagrams.

\begin{figure}[h!]
\includegraphics[width=8cm]{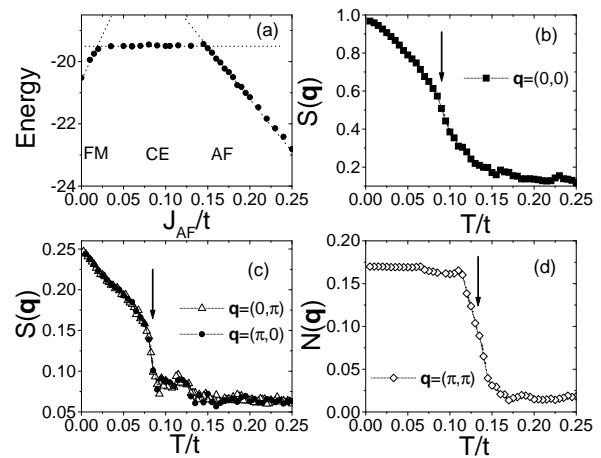}
\caption{(a) MC energy vs. $J_{\rm AF}$ for $x$=1/2, $\lambda $=2.0, $T/t$$\sim$0. 
The location of the level crossings were used in the low-$T$ phase
diagram of Fig.~\ref{fig:figure1}. 
(b) Spin structure factor S(\textbf{q}) as a function of $%
T/t $, for \textbf{q}=(0,0) and $J_{\rm AF}/t$=0, indicating that at low T the
FM phase is stabilized. (c) S(\textbf{q}) vs. $T/t,$ for $J_{\rm AF}/t$=0.05, $%
\lambda $=2.0, and momenta \textbf{q}=(0,$\pi $) and \textbf{q}=($\pi $,0),
characteristic of the CE phase. (d) Charge structure factor N(\textbf{q})
vs. $T/t$ for the same parameters as in (c). In (b), (c), and (d) the
approximate temperatures where correlations become robust are indicated by
arrows. }
\label{fig:figure4}
\end{figure}

Typical results can be found in Fig.~\ref{fig:figure4}. The
level-crossing procedure shown in (a) establishes easily the phase diagram
at low $T$. At finite $T$, the momentum space
correlations show a rapid increase upon cooling at characteristic
temperatures (see Figs.~\ref{fig:figure4}(b,c,d) as typical examples)
due to the increase of the real-space correlations at
the largest available distances. These temperatures are the best approximations
to the true bulk-limit critical temperatures that our investigations can
produce at present.

\begin{figure}[h!]
\includegraphics[width=8cm]{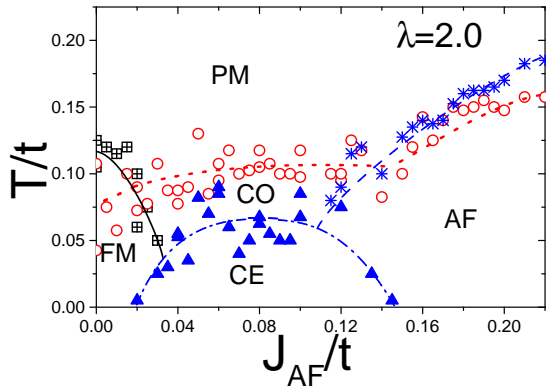}
\caption{Phase diagram $T/t$ vs. $J_{\rm AF}/t$ of the 2D two-orbitals
cooperative-phonon model at $\lambda $=2.0. $T_{\rm C}$, $T_{\rm CE}$, 
and $T_{\rm AF}$
are represented by squares, triangles, and asterisks, respectively. These
temperatures are a crude estimation of the magnetic ordering temperatures of
the FM, CE and AF phases, obtained in our small cluster simulations, as
explained in the text. The oscillations in the results are indicative of the
error bars in the critical temperatures. The temperatures ($T_{\rm CO}$) where
charge and orbital correlations become important upon cooling are marked by
circles and a dotted line. Transitions to the paramagnetic (PM) and CO
phases are second order. Transitions FM-CE, and CE-AF are first-order at low
temperatures. 
}
\label{fig:figure3}
\end{figure}

An example of the phase diagrams constructed by this procedure is given
in Fig.~\ref{fig:figure3}, where the results at $\lambda$=2.0 are shown. 
Here, the `oscillations' in the characteristic temperatures are indicative
of the errors in our procedure, and of the clear 
tendency in simulations at large
$\lambda$ to spend considerable Monte Carlo time trapped in competing states,
as it occurs in glassy systems. In spite of these complications, the
phase diagrams such as Fig.~\ref{fig:figure3} are sufficiently informative
to unveil the dominant properties of the system. 

\begin{figure}[h!]
\includegraphics[width=7cm]{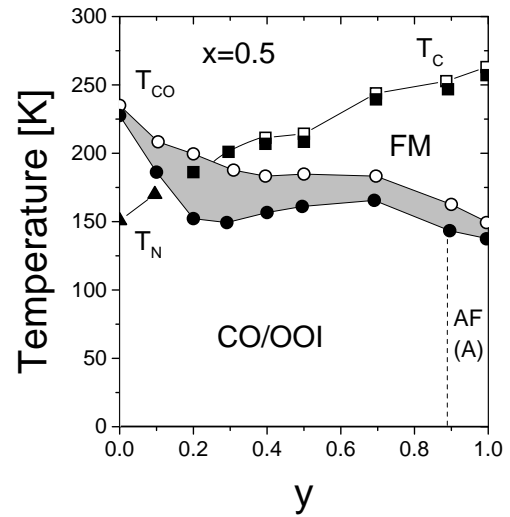}
\caption{Experimental phase diagram of the half-doped compound $%
Pr_{0.5}(Ca_{1-y}Sr_{y})_{0.5}MnO_{3},$ reproduced from Ref~\onlinecite
{Tomioka}. The charge- and orbital-ordered insulator and ferromagnetic
metallic states are denoted by CO-OOI and FM, respectively. The transitions
from (to) the CO-OOI phase are represented by circles. The transitions to
the FM phase are represented by squares. The N\'{e}el temperature is
represented by solid triangles. The grey area indicates hysteresis. 
Note the presence of a FM phase above a CO/AF phase, 
compatible with our results, as discussed in the text.}
\label{fig:figure5}
\end{figure}

Some of the
properties of Fig.~\ref{fig:figure3} are worth explicitly discussing: (i) The N\'eel
critical temperature of the CE state is the lowest among the three dominant
states. This may explain in part its sensitivity to disorder discussed later
in this paper. (ii) At large $\lambda$, charge order in the CE phase 
occurs at a temperature
larger than the N\'eel temperature, as it occurs in many manganite experiments.  
(iii) In the range of $J_{\rm AF}$ between 0.02 and 0.03, the CE phase is
stable at low-$T$, but upon heating the FM state is stabilized, before the
system becomes paramagnetic. This curious behavior is compatible with
experimental results recently reported by Tomioka and Tokura\cite{Tomioka},
reproduced in Fig.~\ref{fig:figure5}. It is remarkable that the model studied
here is able to qualitatively reproduce even this fine detail of the real
phase diagram of $x$=0.5 manganites. The different dimensionalities between
experiments and the simulations reported here are not a problem, since it is
well-known that the phase diagrams of the double-exchange model are qualitatively
similar in the dimensions of interest\cite{book}.

\begin{figure}[h!]
\includegraphics[width=8cm]{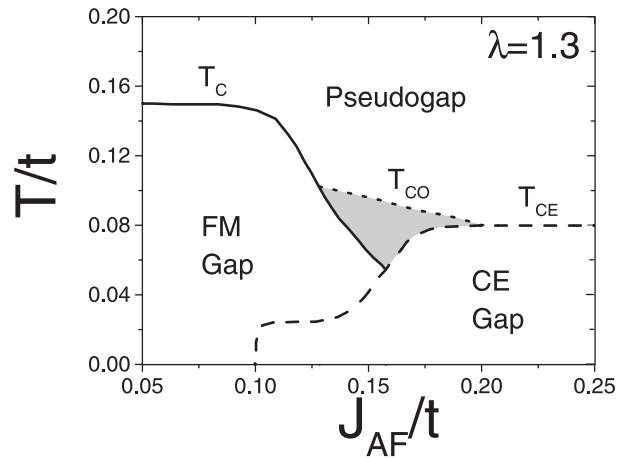}
\caption{Phase diagram of the two-orbitals model with cooperative 
Jahn-Teller phonons at $\lambda$=1.3 and $\beta$=$100$. The results are smooth 
interpolations using the numerically available data, and in this
respect they are not exact. The notation
is standard. Note that at this coupling  there are pseudogap features 
in the density-of-states above the ordering temperatures, and a hard gap
below those critical temperatures. The gray area is a region 
where the simulations
give a mixture of orders, for the large number of iterations carried out.}
\label{fig:figuredavid}
\end{figure}

In Fig.~\ref{fig:figuredavid}, the numerically obtained
phase diagram at $\lambda$=1.3 is also shown, as representative of the
intermediate $\lambda$ range. As in the previous cases,
the information was obtained on a 16-sites lattice, using spin correlations
at the largest available distances to decide which tendency dominates 
in the ground state. Instead of showing explicitly the numbers defining the
lines as done in Fig.~\ref{fig:figure3}, 
here a smooth average is presented for simplicity. Error bars similar
to those of Fig.~\ref{fig:figure3} should be assumed at $\lambda$=1.3 as well.
In good agreement with the results at
other $\lambda$s, the Curie temperature in Fig.~\ref{fig:figuredavid} 
is found to be larger than the CE
critical temperature, and there is a region of couplings where ferromagnetism
is stabilized above the CE state, before turning paramagnetic upon further
heating. The gray region denotes a regime where charge-ordering appears, but
not spin order (however, note that this region with phase competition is
particularly complicated to study due to the presence of many
competing minima in the free energy). 

\begin{figure}[h!]
\includegraphics[width=7cm]{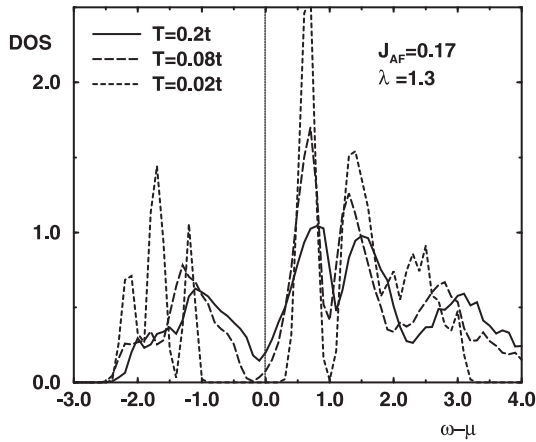}
\caption{Density-of-states obtained numerically at the couplings 
and temperatures indicated.
Comparing with Fig.~\ref{fig:figuredavid}, these results show that pseudogap 
features appear in the DOS above the ordering N\'eel temperature of the CE state, 
at $\lambda$=1.3 and $\beta$=$100$. Pseudogaps have also
been observed in many previous investigations 
(see Refs.~\onlinecite{moreoPG,dessauPG,nohPG}, for instance).}
\label{fig:dospaper}
\end{figure}

Figure~\ref{fig:dospaper} illustrates the presence
of {\it gap} and {\it pseudogap} (PG) features in the density-of-states (DOS), 
due to the robust value
of $\lambda$ used. At low temperatures in the CE phase, there is clearly a gap due
to charge ordering. However, the existence of a reduced DOS at the chemical
potential in the form of a pseudogap survives the increase
of the temperature, and above the N\'eel temperature the effect appears to be caused by 
dynamical Jahn-Teller distortions. This result reinforces
the notion that a large depletion of the DOS at the chemical potential
 should be present in a broad range of couplings and
temperatures in models for manganites. The existence of a PG feature was first
noticed theoretically
 by Moreo {\it et al.} \onlinecite{moreoPG} and experimentally in photoemission
experiments by Dessau {\it et al.} \cite{dessauPG}. 
More recently, in optical conductivity experiments
by Noh and collaborators \cite{nohPG}, PG features were also observed. 
Pseudogaps in DOS alter substantially the transport
properties of the system, and in addition they suggest interesting similarities
with other materials --such as the cuprates-- where pseudogap features have also
been identified. In fact, most of the phenomenology
of the Mn- and Cu-oxides is similar, with phase competition being an important ingredient
to understand phenomenologically their properties \cite{dago-cond-mat}.

\section{Charge Disproportionation}

Our calculations allow us to address theoretically a recent controversy in
the experimental literature related with the actual value of the charge
disproportionation in charge-ordered half-doped manganites. 
The ``standard folklore'' since the 1950s
says that the CE state is made out of a checkerboard distribution 
of 3+ and 4+ charges. However, on theoretical grounds this state appears to be
too ``extreme''. Since the work of Yunoki and 
collaborators\cite{previous} establishing the phase
diagram of manganite models with JT phonons at $x$=0.5, 
it has been noticed that
the CE phase is close to a ferromagnetic metallic phase {\it only at small and
intermediate values of the electron-phonon coupling} $\lambda$. Having the two
phases close to one another is important to understand the phenomenology of
$\rm La_{1-{\it x}} Ca_{\it x} Mn O_3$ (LCMO), 
where it is known that the two phases
``touch'' at $x$=0.5, in the phase diagram varying the Ca concentration.
If indeed $\lambda$ is not very large, then the charge-ordered state should be
far from the 3+/4+ extreme separation. In fact, recent experimental work
by Garcia {\it et al.}, Daoud-Aladine {\it et al.}, and others  have
revealed a very weak signal for charge disproportionation, and the standard
3+/4+ has been severely questioned\cite{no-charge-order}. 

How can we understand this challenging new
phenomenology? A recent key observation by Hotta {\it et al.}\cite{hottaCE}
indicates that the CE state may have an origin totally different from the 
``large Coulomb repulsion'' view of Goodenough (or the large $\lambda$
analog in the presence of JT distortions). In the new perspective, the
formation of the zigzag chains of the CE state
can occur even at very small $\lambda$ since these chains lead to the
 optimization of the kinetic
energy in the presence of a nonzero $J_{\rm AF}$. In this context, 
the insulating character
arises mainly from a band-insulator picture, already present in the study of
individual zigzag chains. The validity of this view is clear in the phase diagram
of Fig.~\ref{fig:figure1}, 
that shows a CE phase that {\it extends to $\lambda$=0}, a result
quite difficult to understand from the large-$\lambda$ standard perspective.
Since the FM metallic and CE phases are not in 
contact at large $\lambda$ in the
theoretical phase diagram, and since experimentally 
they are known to be close in energy (at least for LCMO),
then the 3+/4+ view of the CE state must indeed be revisited.
\begin{figure}[h!]
\includegraphics[width=8cm]{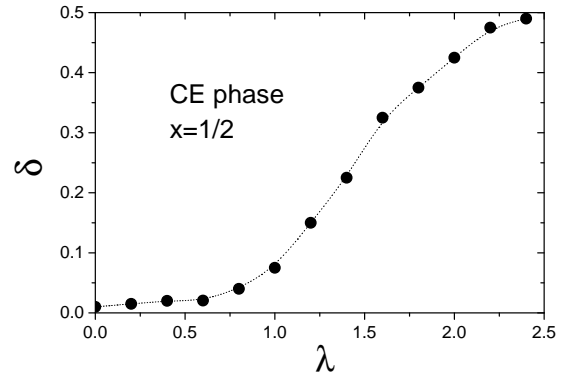}
\caption{Charge disproportionation $\delta $ vs. $\lambda $, calculated
in  the CE phase of Fig.~\ref{fig:figure1}. Starting at $\lambda$=0, the
values of 
$J_{\rm AF}$ used are 
0.25, 0.25, 0.25, 0.25, 0.2, 0.2, 0.2, 0.15, 0.15, 0.1, 0.1, 0.05,and 
0.05, due to the
tilted shape of the CE phase.
A value $\delta \approx $0.5 
corresponds to the standard charge ordering, with Mn$^{3+}$ and Mn$^{4+}$
arranged in a checkerboard pattern. To reproduce this extreme charge-ordered
state an abnormally large value of $\lambda $ is needed. More realistic
parameters correspond to a far less dramatic charge separation.}
\label{fig:figure6}
\end{figure}

Figure~\ref{fig:figure6} 
shows the numerically obtained charge disproportionation  $\delta$
along the CE phase varying $\lambda$
(note that the bended nature of the CE phase in the 
phase diagram Fig.~\ref{fig:figure1} forces
us to adjust $J_{\rm AF}$ as $\lambda$ changes). The entire CE region has a
nonzero $\delta$ according to Fig.~\ref{fig:figure6} -- compatible with its
insulating nature -- 
but clearly there are three qualitatively different
regimes. At small $\lambda$, the small value of 
$\delta$ suggests that the charge 
ordering arises as a consequence of the more dominant
zigzag chain formation, as predicted by Hotta
{\it et al}\cite{hottaCE},
namely the spin order dominates over the charge ordering. Increasing $\lambda$,
a crossover regime is observed in Fig.~\ref{fig:figure6}, 
which is itself followed by the extreme
case of 3+/4+ separation at very large $\lambda$. 
Since the phase diagram of Fig.~\ref{fig:figure1}
shows that the FM charge-disordered and CE phases are in contact only below
$\lambda$=1.2, this implies an upper
bound of approximately $\delta$=0.15 in the charge disproportionation, 
a result far from the more standard assumption $\delta$=0.5.
If $\lambda$ is slightly reduced from 1.2 to 0.8, 
$\delta$ can be as small as 0.05, compatible
with the recent experimental information. 
 As we have stated before, a large Hund coupling suppresses double ocupancy 
of the same orbital. This interaction together with the electron-phonon 
interaction and elastic energies behaves like the Hubbard U interaction\cite{brink}. 
The exact equivalence have not yet been clearly established. Our results are in good 
agreement with van den Brink et al. in the sense that turning on 
$\lambda$ in our study ($U$ in Ref.\onlinecite{brink}), the charge migrates from the 
corner sites of the zig-zag chain, 
to the bridge ones. However, the maximum charge disproportion available in our model 
($\delta$=$0.5$) differs considerably from the $\delta$=$0.185$ result obtained for 
$U$=$\infty$. These discrepancies deserve further investigations to clarify the range 
of validity of both models.
 From all these considerations, 
our conclusion is that indeed the standard extreme view of 
the CE state needs revision. In fact,
the surprising sensitivity of the CE state to the introduction of disorder 
(to be discussed below) may
be indicative of a fragility which intuitively
appears more related to the small-$\lambda$ zigzag-chain-driven 
regime than to the large $\lambda$ limit. Clearly, the last word on the
nature of the CE-state has not been said, and more work should be devoted
to this issue.

\section{Influence of Quenched Disorder
on the Phase Diagram and Fragility of the CE Phase}

\begin{figure}[h!]
\includegraphics[width=8cm]{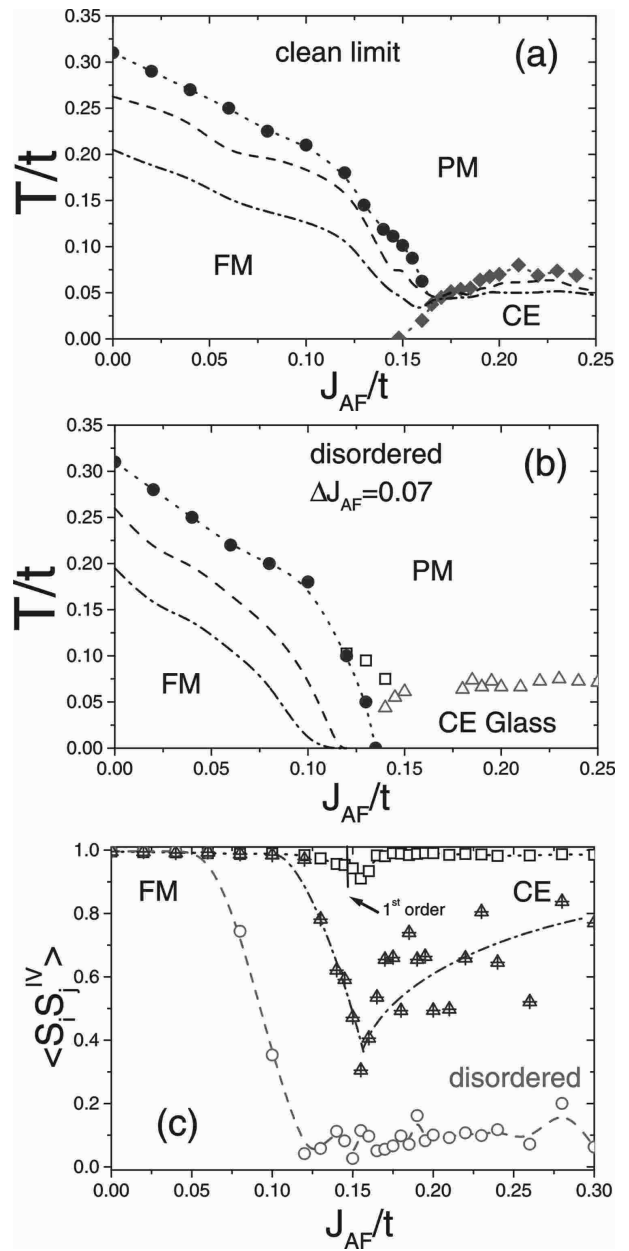}
\caption{(a) Phase diagram $T/t$ vs. $J_{\rm AF}/t$ for the clean limit, with $%
\lambda $=1. The closed circles and diamonds indicate the transition
temperature at which the long-distance correlation $<S_{i}S_{j}^{IV}>$
approximately vanishes upon heating (see text). This is indicative of the
temperature where long-range order develops. Dashed and dash-dotted lines
represent the lines of constant 
$<S_{i}S_{j}^{IV}>$=0.2, and 0.5, respectively. (b) Phase diagram 
$T/t$ vs. $J_{\rm AF}/t$ for the disorder parameter $\Delta J_{\rm AF}$=0.07, and $%
\lambda $=1. The new symbols open -squares and -triangles represent the
transition temperatures to a glassy state with short-range correlations but
with no obvious long-range FM\ and CE order, respectively. The CE state was
found to be very susceptible to disorder. (c) $<S_{i}S_{j}^{IV}>$ vs. $%
J_{\rm AF}/t$ for $\Delta J_{\rm AF}$=0 (squares), $\Delta J_{\rm AF}$=0.04 (triangles),
and $\Delta J_{\rm AF}$=0.07 (circles) at low T. These results are the average 
of five different configurations 
for the bimodal distribution $J_{i}=J_{\rm AF}\pm \Delta J_{\rm AF}$ in the AF links.}
\label{fig:figure7}
\end{figure}

  In Fig.~\ref{fig:figure7}a, the numerically obtained phase diagram at 
fixed $\lambda$=1 varying $J_{\rm AF}$
is shown in the absence of disorder (clean-limit). 
The shape is similar to results presented in 
previous sections at other $\lambda$s, with competing
FM and CE phases. Shown are also lines of constant
spin-spin correlations, which will be useful to understand
the results in the presence of quenched disorder. 
In Fig.~\ref{fig:figure7}b, the modified phase diagram after introducing
disorder in $J_{\rm AF}$ is presented. The disorder is
included by selecting the coupling $J_{\rm AF}$ between localized spins
at every link as 
the uniform value plus an extra contribution $\Delta J_{\rm AF}$, which
is randomly added or subtracted (i.e. a bimodal distribution
is used). In agreement with the expectations
described in the introduction, Fig.~\ref{fig:tg3}, including 
disorder the lines of constant
spin correlations bend and reach zero temperature at particular
values of $J_{\rm AF}$. This is to be contrasted with
the clean-limit case in Fig.~\ref{fig:figure7}a, where the lines of constant correlations
`collide' with the CE phase, rather than 
reaching zero temperature as with disorder incorporated. The
behavior of Fig.~\ref{fig:figure7}b is indicative of the presence of
a spin-disordered window separating the FM and CE phases, as in Fig.~\ref{fig:tg3}c.

However, the
situation appears to be more complicated than previous discussions would have
suggested. The reason is that the CE phase was here found to be abnormally
{\it sensitive to disorder}. At least for the value
of disorder used, a glassy state is obtained
in approximately the same region where the CE phase was found to be stable
in the clean limit. The glassy nature of the CE region appears
in $S({\bf k})$ and it is very clear in
Monte Carlo snapshots (not shown), that present distorted zigzag
chains and patterns of charge ordering -- nearly frozen as the Monte Carlo
time evolves -- in a glassy-looking arrangement. The effect is also manifested in the small
value of correlations, as shown in Fig.~\ref{fig:figure7}c which contains
the spin correlations at the largest distance on the small cluster used.
Without disorder, this correlation has the largest allowed value in both the FM
and CE phases and as the disorder increases, the correlation reduces
its value as expected. What is remarkable is that this 
effect occurs clearly more rapidly in the CE region than
in the ferromagnetic phase. In fact, at $\Delta J_{\rm AF}$=0.07 the spin
correlations are nearly
negligible in the CE regime, indicating the disappearance of order at the
largest distances here available. This is to be contrasted against
the behavior of correlations in the FM phase, which are far more robust. 

The present results are in excellent agreement with the recent experimental
studies of $\rm Ln_{1/2} La_{1/2} Mn O_3$ (Ln = rare-earth) by 
Akahoshi {\it et al.}\cite{akahoshi} and Nakajima {\it et al.}\cite{ueda}.
In these experiments it was observed that the phase
diagram has bicritical behavior (Fig.~\ref{fig:figure7}a) when the crystals are grown
slowly, minimizing by this procedure the influence of disorder. 
However, upon rapid quenching of the growing process to introduce
disorder explicitly, 
it was observed that while the FM phase reduces its critical temperature by a
reasonable amount,
the CE phase is much more affected, turning into a glassy state at low
temperatures. This experimentally-observed abnormal sensitivity of the CE
phase to disorder --here nicely reproduced in the Monte Carlo simulations-- is
expected to be an important ingredient for 
the understanding of the phenomenology of
manganites. Disorder indeed plays a key role in the phase-separated scenario for
Mn oxides\cite{book}, and here its relevance is further confirmed. The simple 
view expressed in Fig.~\ref{fig:tg3}, with a symmetric behavior between the
two competing phases, is more complex in practice.

The disorder 
sensitivity of the CE-phase may arise from the {\it fragility of the zigzag
chains}. As argued before, the realistic regime of couplings, with $\lambda$ 
of order 1, presents a charge-ordering pattern which is far from the
extreme limit 3+/4+, rendering the staggered charge-ordering less robust
than previously believed. In addition, the one-dimensional nature of the
zigzag chains also contributes to this fragility. Imperfections can easily
destroy these nontrivial geometrical arrangement of chains.
Unfortunately, the limited size of the clusters used here does not allow
us to investigate in more detail the observed CE fragility.
These important issues
should be analyzed in more detail in future theoretical and experimental
investigations. It is interesting to remark that 
similar conclusions have been recently reached using
a one-orbital model with cooperative phonons\cite{motome}. Reaching independently
the same conclusion using different models, methods, and lattice sizes
provides confirmation of the fragility of the charge-ordered state that competes
with ferromagnetism. This is an important issue that deserves further studies.

\section{Two Types of CMR}

\begin{figure}[h!]
\includegraphics[width=8cm]{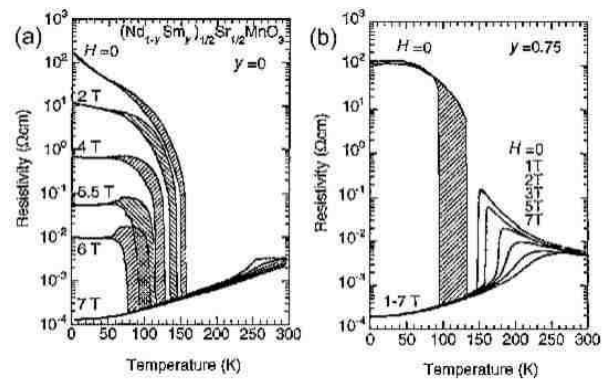}
\caption{Temperature dependence of resistivity under various magnetic fields
for $\rm{(Nd_{1-\it{y}} Sm_{\it{y}})_{1/2} Sr_{1/2} Mn O_3}$
with $y$=0 (a) and 0.75 (b). The hatched area represents thermal hysteresis.
Results from Tokura \textit{et al.}\cite{Tokura}}
\label{fig:tokura}
\end{figure}

Let us start the analysis of d.c. transport properties in the half-doped system
studied here, aiming toward a better understanding 
of the CMR effect. In our opinion,
these effects should actually be divided in two classes: 
(1) the CMR effect that
occurs at low temperatures and corresponds to very abrupt first-order transitions,
and (2) the more standard CMR effect that appears in the regime immediately
above the Curie temperature. Years ago, Tokura and collaborators already
suggested these two types of CMRs\cite{Tokura}. Figure~\ref{fig:tokura} provides 
a direct experimental evidence
of this phenomenon, with a huge effect at low temperatures and a
relatively smaller effect near $T_{\rm Curie}$.

\begin{figure}[h!]
\includegraphics[width=7cm]{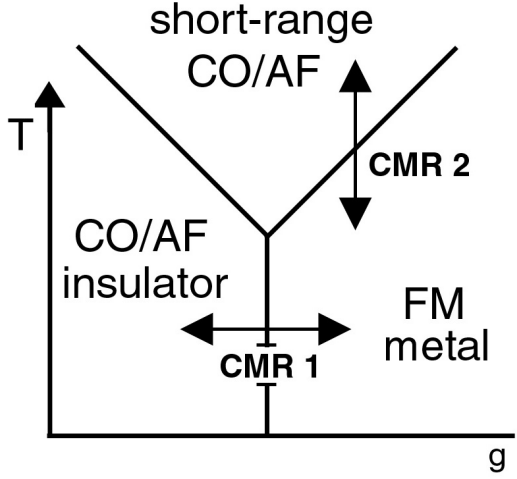}
\caption{ Schematic representation of the generic phase diagram in the
presence of competing FM metal and CO/AF insulator, and for quenched disorder not
sufficiently strong to destroy entirely the first-order transition at low
temperatures. $g$ is a generic variable needed to transfer the system from
one phase to the other. CMR1 and CMR2 are the regions with two types of
large MR transitions, as described in the text (see also Ref.~\onlinecite{burgy}) }
\label{fig:draw35}
\end{figure}

This phenomenology -- with two
CMRs -- has recently been discussed by one of the authors (E.D.) 
in Ref.~\onlinecite{book}, from where Fig.~\ref{fig:draw35} is
reproduced. In this figure, the ``CMR1'' effect corresponds to a direct field-induced
transition from the AF to the FM phases, upon the application of 
an external field. In this case, quenched disorder is $not$ needed
and  the transition occurs even in the clean limit. For CMR1 to occur, 
it is sufficient to be located on the insulating side but
close to the first-order transition. Under these circumstances a 
relatively small field -- which obviously 
favors the FM phase -- renders ferromagnetism
more favorable and an AF to FM first-order jump is induced. On the other hand,
Fig.~\ref{fig:draw35} also shows the  standard ``CMR2'' proccess which is believed to originate
in the more complex percolative regime induced by a nanoscale phase
separated state above the Curie temperature \cite{book,burgy}.  
The realization that there are two types 
of CMRs substantially clarifies
the magnetotransport phenomenology of Mn oxides.

\begin{figure}[h!]
\includegraphics[width=6cm]{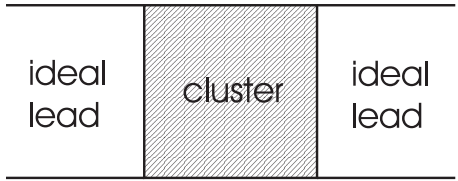}
\caption{Geometrical setup used here for the calculation of the resistance of a
cluster. For more details see Refs.~\onlinecite{verges,alvarez}.}
\label{fig:ideallead}
\end{figure}

To address explicitly  the first type of MR transitions (i.e. CMR1), 
a clean-limit investigation
should be sufficient. To study the resistance of a cluster vs. temperature in the
presence of
magnetic fields, a setup similar to those often used in investigations of mesoscopic
systems will be here employed\cite{verges}. This method was recently applied to
a technically similar problem of spin and carriers in interaction in the context
of diluted magnetic semiconductors\cite{alvarez}.
The interacting 
cluster under investigation is assumed connected
to ``ideal leads'', as shown in Fig.~\ref{fig:ideallead}. 
The information about these leads is included 
through exactly calculated self-energies which are located at the cluster
boundaries. The current circulates after an
infinitesimal voltage drop is included. The technical aspects have been recently
reviewed by Verges\cite{verges} and they will not be repeated here. This technique appears to
be better than other methods often 
used in numerical simulations, that rely on the analysis of the Drude weight $D$
of the optical conductivity with the 
cluster assumed in isolation. In this case, $D$ is often negative due to finite
size effects. In addition, the associated zero-frequency delta function must be
given an arbitrary width to recover a finite resistivity. These problems are
avoided in the present formulation.

\begin{figure}[h!]
\includegraphics[width=8cm]{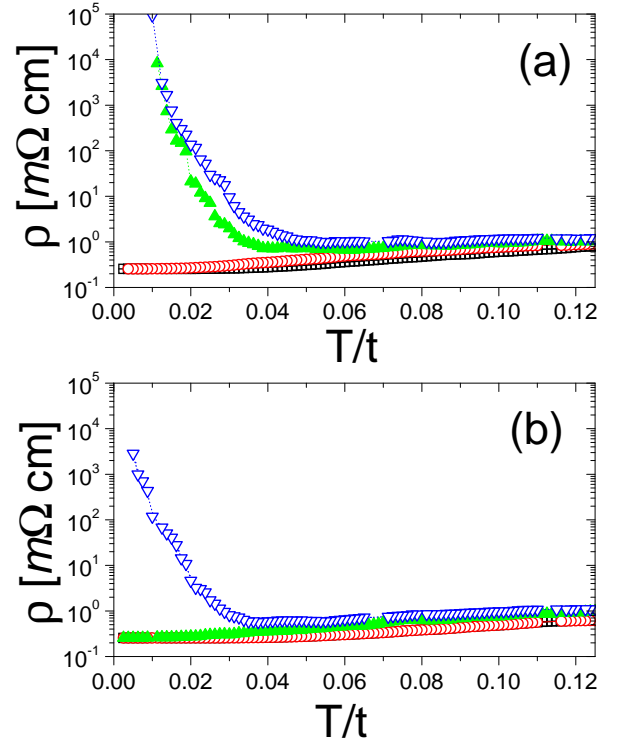}
\caption{Clean-limit investigation of the resistivity $\rho$. Shown is $\rho$ vs.
temperature, at $\Delta J_{\rm AF}$=0, $\lambda $=1, and $J_{\rm AF}$ equal to 0.14
(squares), 0.16 (circles), 0.18 (closed triangles), and 0.2 (open
triangles). The first two $J_{\rm AF}$ cases correspond to the FM regime, and
the other two to the CE state. A first-order transition
separates the metal from the insulator at low temperatures, with concomitant
abrupt changes in the resistance. (a) corresponds to zero magnetic field $%
\mu H/t$=0 while (b) is at $\mu H/t$=0.05. Note in the latter that an insulating 
line in (a), turned metallic in (b), producing a huge MR.}
\label{fig:figure8a}
\end{figure}

A typical result obtained with the setup of Fig.~\ref{fig:ideallead} is shown
in Fig.~\ref{fig:figure8a}. To transform from resistance to resistivity,
a cross section of size 4$a$$\times$1$a$ ($a$=lattice spacing) is assumed.
In the clean limit, and using as reference 
the phase diagram of Fig.~\ref{fig:figure1}, it was observed that
for a range of $J_{\rm AF}$ the material behaves as an insulator since the
CE phase dominates, while for other couplings a metallic behavior is found
once in the FM regime (here insulator and metal are defined simply based on 
the sign of the slope of the resistance vs. temperature curves). 
The technique used here neatly reproduces the expected
behavior of the resistivity vs. temperature for a metal and an insulator. 

\begin{figure}[h!]
\includegraphics[width=8cm]{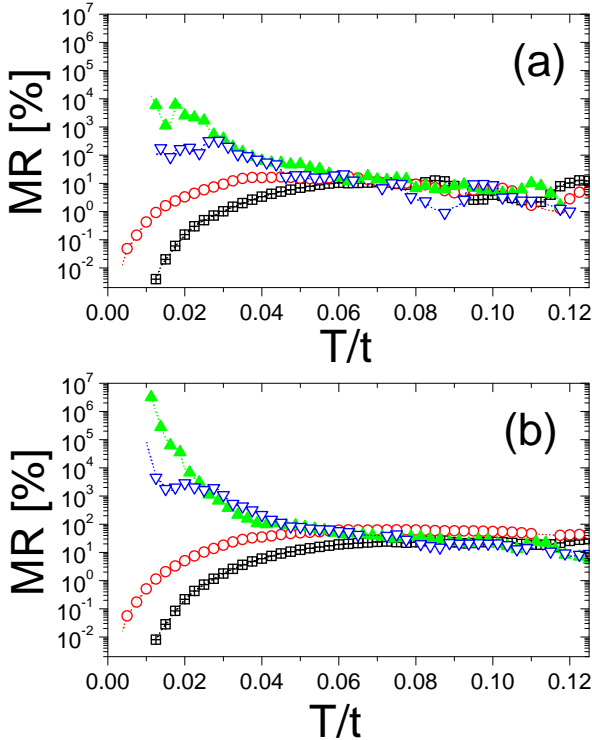}
\caption{Clean-limit magnetoresistance (as defined in text) vs.
temperature considering $\mu H/t$ =0.025 (a), and $\mu H/t$=0.05 (b). The same 
set of parameters and convention of symbols are followed 
as in Fig.~\ref{fig:figure8a}.
Huge MR ratios are observed, as found in experiments.}
\label{fig:figure8b}
\end{figure}

The most important result in this
context occurs when magnetic fields are introduced. In this case, the insulating
behavior found in the CE regime close to the FM phase, more precisely
at  $J_{\rm AF}$=0.18, turns {\it metallic} in a first-order transition
 upon the application of a field of value 0.05$t$ 
(much smaller in magnitude than the natural units
of the problem, such as $t$). Assuming $t$ of the
order of 1,000 K, this is a field of about 50 T, larger but not by a huge amount
with respect to those typically used in CMR experiments. By selecting $J_{\rm AF}$ 
even closer to the first-order transition the value of the field needed to 
induce metallic behavior can be easily reduced.
The associated MR ratios
-- defined as MR=($\rho(0)$-$\rho(H)$)/$\rho(H)$$\times$100$\%$, with $H$ the
field used --
are shown in Fig.~\ref{fig:figure8b}. For the coupling where the insulator-metal transition
was generated by the magnetic field, 
the MR ratio was found to be as large as $10^7 \%$ at low temperatures,
in excellent agreement with the values that can be deduced from the experimental
data in Fig.~\ref{fig:tokura}. As explained
before, this drastic
effect is simply caused by a level crossing induced by the magnetic field,
which favors the FM phase. Note that the $J_{\rm AF}$ coupling used is not
abnormally close to the original transition, namely there is no need to carefully
fix parameters to see this effect. The present calculation clearly shows
that huge MR effects can be obtained in theoretical calculations, even in small
clusters and without the need of tuning couplings. 

\begin{figure}[h!]
\includegraphics[width=8cm]{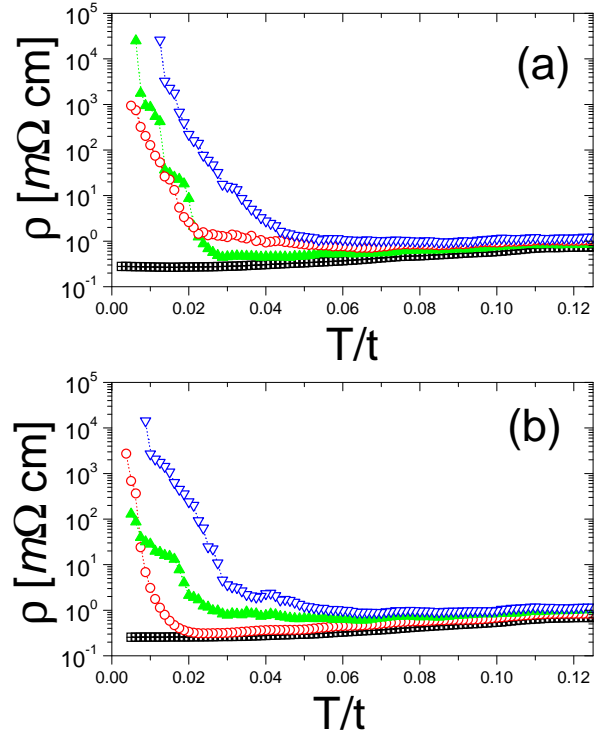}
\caption{Study of the resistivity $\rho$ in the presence of quenched disorder.
Shown is $\rho$ vs. temperature, for $\mu H/t$=0 (a), and $\mu H/t$=0.025 (b),
using $\lambda $=1, $\Delta J_{\rm AF}$=0.04 , and $J_{\rm AF}$ equal to 0
(squares), 0.04 (circles), 0.08 (closed triangles), and 0.12 (open
triangles). Large MR effects are observed, even involving only insulating
phases.}
\label{fig:figure9}
\end{figure}

Regarding the theoretical understanding of the more standard CMR (namely, CMR2
in the present discussion),
previous investigations have relied on sophisticated calculations using simplified
models and resistor networks in order to capture the complex
 percolative physics expected to dominate in real compounds\cite{book,burgy}. 
These subtle effects
cannot be studied in the small clusters currently accessible to
nearly-exact Monte Carlo
studies of realistic models, as those presented here. However,
results in those small clusters when
quenched disorder is included, already provide hints of the physics found
in experiments and in simulations of toy models. 
For example, in Fig.~\ref{fig:figure9}a it is shown that the disorder
reduces the resistivity in the CE regime, and that this effect is further
magnified when magnetic fields are applied. Unfortunately, 
in spite of observing these reasonable
effects, the shape of the resistivity curves still do not show the well-known
metal-insulator transition at a finite temperature. Further computational work 
will be needed to reproduce the CMR2 effect using realistic Hamiltonians.
Fortunately,  the essence of the phenomenon appears to have been captured by
the calculations of Ref.~\onlinecite{burgy}.

\section{Charge-ordering in Electron-doped Materials}

For completeness, here results at electronic densities
$\langle n \rangle$ different from
0.5 are also presented. However, the `half-doped'
character of the investigation remains. To be more specific,
the emphasis in this section is 
on $\langle n \rangle$=1.5, namely in the regime of
{\it electron doping} of undoped compounds such as $\rm La Mn O_3$,
opposite to the hole-doping regime of $\langle n \rangle$=0.5 
discussed in the rest of the paper.
The experimental motivation for this effort relies 
on recent investigations
that have reported results for $\rm La_{0.7} Ce_{0.3} Mn O_3$, 
which is indeed an electron-doped compound having $\rm La Mn O_3$
as the parent material\cite{electron1}. 
A ferromagnetic metallic phase has 
been reported in this context with a Curie temperature $\sim$250~K
and a large magnetoresistance, establishing
clear similarities with the well-known results for the hole-doped
region of the phase diagram.

These experiments suggest that there
must be a relation between electron and hole doping, which has not
been discussed theoretically before to our knowledge.
In particular, it is interesting to speculate what kind of states
will be obtained in the electron-doped regime once the doping is
made as large as 50\% (i.e. $\langle n \rangle$=1.5). In other words,
here the study will focus on the discussion of what kind of CO/OO state
could be obtained in the {\it electron half-doped} regime of manganites,
justifying the inclusion of the results described below in the present
paper, which is devoted to half-doped manganites in general.

\begin{figure}[h!]
\includegraphics[width=6cm]{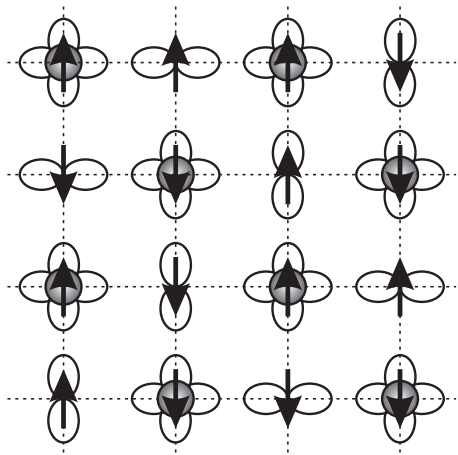}
\caption{Orbital, spin, and charge arrangement for the electron 
overdoped ($\langle n \rangle$=1.5) 
CE/CO/OO phase. The $3x^{2}-r^{2}$ and $3y^{2}-r^{2}$ 
orbitals represent the Mn$^{3+}$ ions. The superposition of the
$x^{2}-y^{2}$ plus $3z^{2}-r^{2}$ orbitals
(closed circles) represents the Mn$^{2+}$ ions.}
\label{fig:18}
\end{figure}

The main conclusion of the analysis of the $\langle n \rangle$=1.5
phase diagram using Monte Carlo simulations
is that in the large Hund coupling limit, there is
a mapping between $\langle n \rangle$ bigger and smaller than 
one (with $\langle n \rangle$=1 corresponding to $\rm LaMnO_3$). 
In fact, in the $J_{\rm H}$=$\infty$ limit studied here
the phase diagrams in the plane $\lambda$-$J_{\rm AF}$
are {\it identical} (within numerical accuracy)
with regards to the location of the phase boundaries. However,
the characteristics of each phase cannot be the same in
view of the different electronic densities. As particular case,
our effort predicts that the analog of the well-known
CE phase of half hole-doped manganites is the state shown schematically
in Fig.~\ref{fig:18}. 
In this state, 
there is a separation of charge into a `3+/2+' like configuration,
to be contrasted with the standard 3+/4+ state at $\langle n \rangle$=0.5.
The zigzag chains of the standard CE-state are still present, and even
the orbitals in the middle of the segments of the zigzag chains
are elongated along the $x$ and $y$ axis. The main difference between
$\langle n \rangle$=0.5 and 1.5 is the state at the vertices of those 
chains, which is empty at density 0.5 but it contains two electrons at 1.5.
An analog situation is found in the other charge-ordered phases.
It would be very interesting to confirm experimentally
whether this state is present in real half-electron-doped
manganites. Note that our calculation
does not include the effect of Coulombic interactions, which may be more
important in the electron-doping regime than in the hole-doping region.
However, in other materials such as the cuprates, a strong similarity
between electron- and hole-doping has been found as well, 
suggesting that a similar situation may occur in manganites. Completing
the phase diagram of electron-doped manganites would help us in achieving
a deeper understanding of these materials. The concrete prediction of
our efforts is that there should be an approximate particle-hole
symmetry with respect to the undoped $\rm La Mn O_3$ compound, in a
similar spirit as it occurs in the cuprates\cite{comment-ph}. 

\section{Conclusions} 

The investigations reported in this paper have unveiled several unexpected
properties of half-doped manganites. 
For example, the CE phase was found to be more
sensitive than expected to the addition of quenched disorder. The effort
in this context included the calculation of the clean-limit phase diagram 
in the $T$-$J_{\rm AF}$ plane, to understand the competition between
the FM and CE states believed by most manganite experts 
to be at the heart of the CMR phenomenon. 
In agreement with previous calculations\cite{book,burgy}, a
low-temperature first-order transition between the two phases was found,
with a shape similar to those reported experimentally\cite{Tomioka}.
Adding disorder, the predicted\cite{burgy} depletion of critical temperatures was
observed, but contrary to those expectations the reduction was not symmetrical 
in magnitude for the two competing phases. In fact, it was observed that
the CE state
rapidly transforms into a ``CE glass'' with disorder, while the ferromagnetic
phase is comparatively less affected, in nice agreement 
with recent experiments where disorder is introduced ``by hand'' upon a
clean-limit bicritical-shaped phase diagram\cite{akahoshi,ueda}.
The zigzag chains of the
CE state and its very subtle arrangement of spins, charge, and orbitals
seems easily broken by imperfections, contrary to the more robust 
uniform FM state. 

The present results revealed other new properties of
manganite states that also deserve further 
investigations. Among these interesting
new properties are the existence of {\it novel phases}, such as
a FM/CO state,
and a discussion on charge disproportionation that suggests
that the widely held view of charge-ordering as containing 3+ and 4+ ions
needs considerable revision. 

A second important result presented here is the existence of a very large
MR effect when the insulating state is close in energy to the FM state,
as it occurs in the first-order transitions systematically found in the
clean limit. As a consequence, without quenched disorder and working
at low temperatures, the insulating state can be destabilized by the
FM state with increasing fields. The magnitude of the field required
can be very small in units of the hopping, and its actual critical
value depends on how close the CE state investigated 
is to the FM state in the phase diagram (i.e. how close the analyzed
value of $J_{\rm AF}$ is to the critical value separating
the phases). This MR effect is {\it qualitatively different} from the more
standard effect found above the Curie temperature, which is believed
to need disorder effects to be understood\cite{book,burgy}. As predicted
by Tokura and collaborators\cite{Tokura} many years ago, {\it there are
two types of CMR}: a low-temperature form -- investigated theoretically
here -- and a higher temperature variety with the standard profiles for
resistivity vs. temperature. In carrying out these investigations, we
have introduced techniques borrowed from mesoscopic physics, within the
context of the Landauer formalism.

The study of electron-doped systems has also been initiated 
in the investigations  reported in this paper,
focusing on the half-doped limit. Charge-ordered states analogous to those
observed at half-hole-doping were identified. A formal particle-hole 
symmetry appears to exist in the system with respect to the undoped 
$\rm LaMnO_3$ limit, at least  for a  large Hund coupling.
More theoretical work in this area should be pursued, since 
experiments are starting to investigate electron-doped manganites.

It is clear that the study of models for manganites using unbiased techniques
provides a plethora of interesting results.
In the field of manganites the 
crossfertilization theory-experiments has been remarkably positive,
and new surprises will likely be found in the near future. Investigations
of manganites and other related oxides should continue at its present
rapid rate, to enhance our understanding of correlated electron systems
in general.

\section{Acknowledgments}
H. A. is supported by the Schuler Fellowship at the Magnet Lab. 
A. M. and E. D. are supported by the NSF grant DMR-0122523. Additional
funds have been provided by Martech (FSU). 
We acknowledge the help of G. Alvarez and J. Verges
in the study of the resistance of clusters reported here and 
useful conversations with T. Hotta and L. Brey. The School of Computational Science 
and Information Technology (CSIT) at FSU is also acknowledged.
 
\section{Appendix: Simulations in the Canonical Ensemble}
In most of our simulations, we have kept the number of particles fixed in
our systems basically at each Monte Carlo Step per Site (MCS/S). In order to
achieve that, we have to solve the equation,
$$
n-\sum_{i=1}^{N}\sum_{k=1}^{2N}\frac{c_{k,i}^{+}c_{k,i}}{1+\exp [
(E_{k}-\mu)/T ]}=0
$$
\noindent
for the chemical potential, $\mu $; where $n$ is the desired number of
electrons in the system, $N$ is the number of sites of the cluster, $%
c_{k,i}^{+}$ and $c_{k,i}$ are the fermionic operators after diagonalizing
the hamiltonian matrix (Eq. (1)), and $E_{k}$ are the 
electronic energy levels. These energies are ordered in the way: $E_{2N}\leq
E_{k}\leq E_{1}$, with $k$ running between 1 and $2N$. The first sum symbol
runs over sites, while the second runs over energy levels ($2N$ in our case,
since we have two orbitals per site).

We have solved this equation using the Newton-Raphson method \cite{numrec},
starting with an initial seed $\mu _{0}=1/2(E_{2N-l}+E_{2N-l+1})$, with $l=$%
int$(xN)$, and the filling $x$ given by: $x=n/N$. The symbol int( ) means
the nearest integer of a real number.
Fixing an absolute error for $n$ of $10^{-5}$, the chemical potential $\mu $
is typically found in 4-5 iterations. At sufficiently low temperatures, 
$k_{\rm B}T/t\leq$0.01 for example, the convergence of this method can fail. In
this case we use the auxiliary bisection method \cite{numrec}, where typically 
20-30 steps are needed for the same accuracy.

\end{document}